\documentstyle[12pt,procsla,epsfig]{article}

  \catcode`\@=11
  \long\def\@makefntext#1{
  \protect\noindent \hbox to 3.2pt {\hskip-.9pt
  $^{{\ninerm\@thefnmark}}$\hfil}#1\hfill}              %CAN BE USED

  \def\@makefnmark{\hbox to 0pt{$^{\@thefnmark}$\hss}}  %ORIGINAL

  \def\ps@myheadings{\let\@mkboth\@gobbletwo
  \def\@oddhead{\hbox{}
  \rightmark\hfil\ninerm\thepage}
  \def\@oddfoot{}\def\@evenhead{\ninerm\thepage\hfil
  \leftmark\hbox{}}\def\@evenfoot{}
  \def\sectionmark##1{}\def\subsectionmark##1{}}

\begin{document}

  \centerline{\normalsize\bf THE LAKE BAIKAL NEUTRINO EXPERIMENT: SELECTED RESULTS}
  \baselineskip=16pt

  \vfill
  \vspace*{0.6cm}
%%\centerline
\noindent
{\footnotesize
{V.A.BALKANOV$^a$, I.A.BELOLAPTIKOV$^g$, L.B.BEZRUKOV$^a$, N.M.BUDNEV$^b$, A.G.CHENSKY$^b$,
I.A.DANILCHENKO$^a$, Zh.-A.M.DZHILKIBAEV$^a$, G.V.DOMOGATSKY$^a$, A.A.DOROSHENKO$^a$, 
S.V.FIALKOVSKY$^d$, O.N.GAPONENKO$^a$, A.A.GARUS$^a$, T.I.GRESS$^b$, D.KISS$^i$,
A.I.KLIMOV$^f$, S.I.KLIMUSHIN$^a$, A.P.KOSHECHKIN$^a$, Vy.E.KUZNETZOV$^a$, V.F.KULEPOV$^d$, 
L.A.KUZMICHEV$^c$, S.V.LOVZOV$^b$, J.J.LAUDENSKAITE$^b$, B.K.LUBSANDORZHIEV$^a$,
M.B.MILENIN$^d$, R.R.MIRGAZOV$^b$, N.I.MOSEIKO$^c$, V.A.NETIKOV$^a$, E.A.OSIPOVA$^c$,
A.I.PANFILOV$^a$, Yu.V.PARFENOV$^b$, A.A.PAVLOV$^b$, E.N.PLISKOVSKY$^a$, P.G.POKHIL$^a$,
E.G.POPOVA$^c$, M.I.ROZANOV$^e$, I.A.SOKALSKI$^a$, CH.SPIERING$^h$, O.STREICHER$^h$,
B.A.TARASHANSKY$^b$, G.TOHT$^i$, T.THON$^h$, R.VASILIEV$^a$, R.WISCHNEWSKI$^h$,
I.V.YASHIN$^c$.}}

\vspace{3mm}
%\noindent
{\footnotesize\it
$^a$\ Institute for Nuclear Research, Moscow, Russia

$^b$\ Irkutsk State University, Irkutsk, Russia

$^c$\ Institute of Nuclear Physics, MSU, Moscow, Russia

$^d$\ Nizhni  Novgorod  State  Technical University, Nizhni  Novgorod, Russia

$^e$\ St.Petersburg State  Marine Technical  University, St.Petersburg, Russia

$^f$\  Kurchatov Institute, Moscow, Russia

$^g$\ Joint Institute for Nuclear Research, Dubna, Russia

$^h$\ DESY-Zeuthen, Berlin/Zeuthen, Germany

$^i$\ KFKI, Budapest, Hungary
}

  \centerline{\footnotesize presented by G.V.Domogatsky }
  \centerline{\footnotesize {\it Institute for Nuclear Research, Moscow, Russia}}
  \centerline{\footnotesize E-mail: domogats@pcbai10.inr.ruhep.ru}

\vspace*{0.9cm}
\abstracts{
We  review  the  present  status  of  the  lake  Baikal
Neutrino  Experiment  and  present   selected   physics
 results gained with the consequetive stages of the  stepwise
 increasing detector: from  NT-36  to  NT-96.  Results  cover
 atmospheric muons,  neutrino  events, very high energy neutrinos,
 search  for  neutrino
 events  from  WIMP  annihilation,   search   for   magnetic
 monopoles and environmental studies. We  also  describe an air
 Cherenkov  array  developed  for  the   study   of   angular
 resolution of NT-200.
}
  %\vspace*{0.6cm}
  \normalsize\baselineskip=15pt
  \setcounter{footnote}{0}
  \renewcommand{\thefootnote}{\alph{footnote}}
  \section{Introduction }

      The Lake Baikal Neutrino Experiment exploits  the  deep
 water of the great Siberian lake as a detection  medium  for
 high energy neutrinos via muons and electrons, generated  in
 high energy neutrino interactions. High energy neutrinos are
 of particular importance for high  energy  astrophysics  for
 the last few decades to shed light on the physics of  Active
 Galactic Nuclei (AGN), binary star systems, gamma ray  burst
 (GRB) etc.
      The life time of the lake  Baikal  Neutrino  Experiment
 spans almost two decades, from first small experiment  with
 a few PMTs to the present  large  scale  neutrino  telescope
 NT-200 \cite{Project,APP,APP2},
 which has been put into full operation on  April  6th
 1998. The effective area of the telescope for muons is 2000-10000 m$^2$
depending on the muon energy. The expected rate of muons from
atmospheric neutrinos, with a muon energy thresholds of 10 GeV
and after all cuts rejecting background, is about 1 per two days.

   \section{Neutrino Telescope NT-200 }
   \subsection{Detector and Site}
      The Neutrino Telescope NT-200 is  located  in  the
 southern part of Lake Baikal  (51.50N  and  104.20E)  at
 3,6km  from  the  shore  and  at  a  depth  of  ~1km.  The
 absorption length $L_{abs}$ of water at the site for  wavelengths
 between 470nm and 500nm is about 20m and seasonal variations
 are less than 20\%. The light  scattering  is  subjected
 strongly to seasonal variations and from year to year. We  can
 say now that light scattering is rather strongly anisotropic
 and typical values of $L_{scatt}$ are about 15m.
      NT-200 (fig.1) consists of  192  optical  modules  (OMs)  at  8
 strings arranged at an umbrella like  frame \cite{Project,APP}.  Pairs of
 OMs are switched in  coincidence  with  15ns
 time window  and define a  channel. We  pursue  pairwise
 ideology  for  many  reasons:  to  suppress  individual   OM
 background counting rates due to OM dark current and water
 luminescence, the level of late- and afterpulses etc.
      The  OMs \cite{NIM}  consists  of  QUASAR-370   phototube \cite{Aachen,Pylos}
 enclosed in transparent, nearly spherical pressure  housing.
 The optical contact between the photocathode region  of  the
 phototube  and  the  pressure  sphere  is  made  by   liquid
 glycerine sealed with a layer of  polyurethane.  Apart  from
 the phototube every OM contains two HV power supplies  (\mbox{25 kV}
 and \mbox{2 kV}),  a  voltage   divider,   two   preamplifiers,   a
 calibration LED and a vacuum probe.
      The QUASAR-370 phototube has been developed  especially
 for the Lake Baikal Neutrino Experiment  by  INR  and  KATOD
 Company in Novosibirsk. The phototube is a hybrid  one  and  has  excellent
 time and amplitude resolutions.

      The detector electronics system \cite{APP} is hierarchical:
 from  the  lowest  level  to  the  highest   one   --   OM's
 electronics,   "sviazka"   electronics   module,   string
 electronics module  and  the  detector  electronics  module,
 where  detector  trigger  signals   are   formed   and   all
 information from the string electronics module  are received  and
 sent to the shore station. The detector is operated from the
 shore station.
      A muon trigger is formed by the requirement of \mbox{$\geq N$ hits}
 (with hit  referring  to  a channel)  within  \mbox{500 ns}.  $N$  is
 typically set  to  the  value  \mbox{3 or 4}.  For  such  events,
 amplitude and time of all fired channels are  digitized  and
 sent to the shore. The event records includes all hits within  a
 time window of \mbox{-1000 ns to +800 ns}  with  respect  to  the
 muon trigger signal.
      A  separate  monopole  trigger  system   searches   for
 clusters of sequential hits in individual channels which are
 characteristic for the  passage  of  slowly  moving,  bright
 objects like GUT monopoles.

      There  is a separate  hydrological   string   to   study
 permanently water parameters of the  lake.  This  string  is
 deployed at about 60m from the main part of NT-200.

      There are two nitrogen lasers  for  the  detector  time
 calibration placed just above and below  the  detector.  The
 former one illuminates each  individual  channel  via  fiber
 optics and the latter illuminates the detector as a whole.

\subsection {Cherenkov EAS Array}

      To determine angular resolution of  the NT-200  mobile
 wide angle air cherenkov  array  has  been  developed \cite{Calg}.  This
 array (Fig.2) has been deployed for the last two expeditions on  the
 ice just above NT-200. It consists of four  optical  modules
 put on the sledges for fast deployment. Three  of  them  are
 fixed in the vertices of an equilateral triangle
 and another one  just  in  the  center  of  the
 triangle. The sidelength of the triangle  is  about
 \mbox{100 m}. The analog signals from the optical  modules  are  fed  by
 electrical coax cables to  the  central  electronic  station
 which is located near the central optical module.  The  data
 acquisition   system    includes    4    constant    fraction
 discriminators, a majority coincidence  unit,  two  TDCs  with
 \mbox{500 ns} and \mbox{5000 ns} ranges, ADC, an EAS event counter, counting rate
 scalers for  each  channel  and an underwater  master  signals
 counter.

      Each   optical    module    incorporates   QUASAR-370G
 phototubes, two high voltage power supplies of \mbox{25 kV} and \mbox{1 kV}
 for  an  electron-optical   preamplifier   and   small   PMT
 respectively,  anode  pulses  preamplifier   and   LED   for
 amplitude calibration. The phototube is arranged into light tight
 metallic box which is equipped with mechanically removable lid.
 To  increase the sensitivity  area,  Winston  cones  are   used
 providing almost \mbox{2450 cm$^2$} final sensitive area by a
 factor of 2 larger than in the 1998 array. The angular
 acceptance of the optical module is restricted to 30$^0$  of  half
 angle. The QUASAR-370G phototube is practically the same  as
 the  phototube used in NT-200 but with  six  stages  small  PMT
 developed to withstand high mean anode current due  to  night
 sky background (NSB). To stabilize the phototubes gain, the  HV  power
 supplies are surrounded by a thermostat.

      A 4-fold coincidence within a 1000 ns gate  defines the trigger  of
 the array. Using light concentrators allows to increase  the
 array trigger rate by a factor  of  two.  The  trigger  rate
 depends on the weather conditions and is in  average  to
 0.8-1Hz.
      The trigger signal of underwater telescope is  fed  via
 more than 1km long  coax  cable  to  the  center  electronic
 station of the Cherenkov array and switched into coincidence
 with the array trigger signal. The time difference between  them
 is measured by a wide range TDC. The  synchronization  between
 Cherenkov array and underwater  telescope  is  done  by
 comparing two  underwater  event  counters  in  the  central
 electronic station on the ice and detector electronic module
 of the underwater telescope read out from the shore station.
      The array energy threshold is about \mbox{100 TeV}.  The
 angular resolution of the array  is $\sim0.5^{\circ}\div1^{\circ}$  giving  a
 good reference  point  to  estimate the  ngular  resolution  of the 
 underwater  telescope  for  muons  close  to  the   vertical
 direction ($0^{\circ}\div30^{\circ}$) since high energy  muons  retain  the
 direction of their parent shower.
      In 1998 the EAS array operated in coincidence with only one
 string, and overall 450 events were  registered.  Analysis  of
 those  events  showed  that  the  accuracy  of  zenith  angle
 reconstruction taking into account only time information  is
 close to  $6^{\circ}$,  what  is  in  reasonable   agreement   with   MC
 calculations ($5^{\circ}$). In 1999 we  managed  to  establish the  joint
 work of EAS array and NT-200, but unfortunately we collected   only
 150 events due to bad weather.  The
 analysis of data is still in progress and  results  will  be
 presented later.

\subsection {Shadow of the shore in muons}

      Muon angular distributions as well as depth  dependence
 of the vertical flux obtained from  data  taken  with  NT-36
 have been presented earlier \cite{APP}.  Another  example  which
 confirms the efficiency of  track  reconstruction uses
 the  shore  "shadow"  in
 muons recorded with NT-96.

      As it was already mentioned, NT-200  is  situated  at  a
 distance of 3.6 km to the nearby shore of the lake, and  more
 than \mbox{30 km} to the opposite shore.
 This asymmetry allows to study the asymmetry in the
 azimuth distribution of muons  under  large  zenith  angles,
 where reconstruction for the rather  "thin"  NT-96  is  most
 critical.

      A sharp decrease of the muon intensity at zenith angles
 of  $70^{\circ}\div90^{\circ}$  is  expected.  The  comparison   of   the
 experimental muon angular distribution with MC  calculations
 gives us an estimation of the accuracy of the reconstruction
 error close to the horizontal direction.

      Indeed, the NT-96 data show clear dip  of
 the muon flux in the direction of the shore and  for  zenith
 angles larger than $70^{\circ}$ (fig.3) -- in  very  good  agreement  with
 calculations which take into consideration the effect of the
 "shadowing" shore.

  \section{Selected physics results}
In this chapter we present some physics results based on the data of 70 days
effective operating time of the 4-string array NT-96.

\subsection{Separation of neutrino events with full track reconstruction}
The signature of neutrino induced events is a muon crossing
the detector from below.
With the flux of downward muons exceeding that of
upward muons from atmospheric neutrino interactions by
 about 6 orders of magnitude, a careful reconstruction is of
prime importance.
The reconstruction algorithm \cite{APP} is based the assumption
that the light induced by muons is emitted exactly under the Cherenkov
 angle ($42^{\circ}$) relative to the muon track.
For full track reconstruction ($\theta$, $\phi$ and spatial coordinates)
one needs more than 5 hit channels on more than 3 strings.
In contrast to first stages of the detector
(NT-36 \cite{FRST_vert}),
NT-96 can be
considered as a real  neutrino telescope for a wide region in
zenith angle $\theta$.
After the reconstruction of all events with $\ge$ 9
hits at $\ge$
3 strings (trigger 9/3), quality cuts have been applied
in order to reject fake events.
Furthermore, in order to guarantee a minimum lever arm for track
fitting, events with a projection of the most distant channels on
the track ($Z_{dist}$) less than 35 meters have been rejected.
Due to the small transversal
dimensions of NT-96, this cut excludes zenith angles close to
the horizon.

The efficiency of the procedure has been tested with
a sample of $ 1.8 \times 10^6$ MC-generated atmospheric
muons, and with MC-generated upward muons due to atmospheric neutrinos.
It turns out that the signal to noise ratio is $ > 1$ for this sample.

The reconstructed angular distribution of \mbox{$2 \times 10^7$} events
taken with
NT-96 in April/September 1996 -- after all cuts -- is
shown in Fig.4.
%
%\vspace{-7mm}
%

From 70 days of \mbox{NT-96} data,
12 neutrino candidates have been found. Nine of them have been
fully reconstructed. Three nearly upward vertical tracks
(see subsection 3.2)
hit only 2 strings and give a clear zenith angle but
ambiguities in the azimuth angle -- similar to the two events from
NT-36 \cite{APP}.
This is in  good agreement with MC expectations.

\subsection{Search for nearly vertical upward moving neutrinos}
Unlike the standard analysis \cite{APP}, the method presented
in this section relies on the application of
a series of cuts which are tailored to the response
of the telescope to nearly vertically upward moving muons \cite{FRST_vert,INR_vert}.
The cuts remove muon events far away from the opposite
zenith as well as  background events which are mostly due to
pair and bremsstrahlung showers below the array and to naked downward
moving atmospheric muons with zenith angles close to the horizon
($\theta>60^{\circ}$).
The candidates identified by the cuts are afterwards fitted in order to
determine the zenith angle.
We included all events with $\ge$4 hits along at least one
of all hit strings.
To this sample, a series of 6 cuts is applied. Firstly,
the time differences of hit channels along each individual  string
have to be compatible with a particle close to the opposite zenith(1).
The event length should be large enough(2), the maximum recorded
amplitude  should not exceed a certain value(3), and
the center of gravity of hit channels should not be close to
the detector bottom(4). The latter two cuts reject
efficiently brems showers from downward muons.
Finally, also time differences of hits along different
strings have to correspond to a nearly vertical muon (5) and
the time difference between top and bottom hit in an event
has to be larger than a minimum value (6).

The effective area for muons moving close to opposite zenith
and fulfilling all cuts  exceeds $1000$ m$^2$.

Within 70 days of effective data taking, $8.4 \times 10^7$ events
with the muon trigger $N_{hit} \ge 4$ have been selected.

Table\,1  summarizes the number of events
from all 3 event samples (MC signal and background, and experiment)
which survive the subsequent cuts.
After applying all cuts, four events were selected as neutrino
candidates, compared to 3.5 expected from MC.
One of the four events
has 19 hit channels on four strings and was selected
as neutrino candidate by the standard analysis too. The zenith angular
distribution of these four neutrino candidates
is shown in the inner box of Fig.3.

Regarding the  four detected events as being due to
atmospheric neutrinos, one can derive an
upper limit on the flux of muons from the center of the Earth
due to annihilation of neutralinos - the favored candidate for
cold dark matter.

The limits on the excess muon flux obtained with underground
experiments \cite{Bak,MACRO,Kam} and NT-96 are shown in Table 2.
The limits obtained with NT-96
are 4--7 times worse then the best underground limits since
only the first 70 days of NT-96 have been analysed.
%
%\vspace{-4mm}

This result, however, illustrates
the capability of underwater experiments with respect
to the search for muons due to
neutralino annihilation in the center of the Earth.
MC shows that for NT-200 the effective area is about 2000 m$^2$, 
for $E_{\mu}$$>$10GeV, two times larger than for NT-96.
In case that the energy threshold for upward muons could be decreased to
5 GeV NT-200 will permit to select a non-negligible amount
of contained events and estimate the energy of muons. This will allow
to study the neutrino energy spectrum  for neutrinos having crossed
about \mbox{13000 km} in the Earth \cite{Mosc}. Estimates show that
the full number of contained events will be about 20 for $\theta>165^{\circ}$
per year. In case of $\nu_{\mu}$--$\nu_{x}$ oscillations, the $\nu_{\mu}$
flux will be suppressed and for $\Delta{m}^{2}$=$10^{-3}$eV$^{2}$ we will
find only 7 events.

\subsection{High Energy Neutrinos}

The ultimate goal of large underwater neutrino telescopes
is the identification of extraterrestrial high energy neutrinos.
In this chapter we present results of a search for
neutrinos with $E_{\nu}>10 \,$TeV derived from NT-96 data.
Cherenkov
light emitted by the electro-magnetic and (or) hadronic
particle cascades and high energy muons
produced at the neutrino interaction
vertex in a large volume around the neutrino telescope.
Earlier, a similar strategy has been used by the \mbox{DUMAND
\cite{DUMAND}} and the \mbox{AMANDA \cite{AMANDA}}
collaborations to obtain upper limits on the diffuse
flux of high energy neutrinos.

Fig.5 illustrates the detection principle. We select
events with high multiplicity of hit channels
corresponding to bright cascades. The
volume considered for generation of cascades is essentially
below the geometrical volume of NT-96 -- its upper plane
crosses the center of the telescope. A cut is applied
which accepts only time patterns corresponding to upward
traveling light signals (see below). This cut rejects most events
from brems-cascades along downward muons since the majority
of muons is close to the vertical; they would
cross the detector and
generate a downward time pattern. Only the
fewer muons with large zenith
angles may escape detection and illuminate the array
exclusively via bright cascades below the detector.
These events then have to be rejected by a stringent
multiplicity cut.

Neutrinos produce showers and high energy muons through
CC-interactions

\begin{equation}
\nu_l(\bar{\nu_l}) + N \stackrel{CC}{\longrightarrow} l^-(l^+) +
\mbox{hadrons},
\end{equation}
through NC-interactions

\begin{equation}
\nu_l(\bar{\nu_l}) + N \stackrel{NC}{\longrightarrow}
\nu_l(\bar{\nu_l}) + \mbox{hadrons},
\end{equation}
where $l=e$ or $\mu$, and through resonance production \cite{Glash,Ber1,Gandi}

\begin{equation}
\bar{\nu_e} + e^- \rightarrow W^- \rightarrow \mbox{anything},
\end{equation}
%\newpage
\noindent
with the resonant neutrino energy
$E_0=M^{2}_w/2m_e=6.3\times 10^6 \,$GeV
and cross section $5.02\times 10^{-31}$cm$^2$.

Within the first 70 days of effective data taking, $8.4 \times 10^7$ events
with the muon trigger $N_{hit} \ge 4$ have been selected.

For this analysis we used events with $\ge$4 hits along at least one
of all hit strings. The time difference between any two channels
deployed on the same string was required to obey the condition:

\begin{equation}
\mid(t_i-t_j)-z_{ij}/c\mid<a\cdot z_{ij} + 2\delta, \,\,\, (i<j).
\end{equation}
The $t_i, \, t_j$ are the arrival times at channels $i,j$, and
$z_{ij}$ is their vertical distance. With $\delta=5$nsec
accounting for the timing error, the condition
$\mid(t_i-t_j)-z_{ij}/c\mid <2\delta$ (i.e. $a=0$)
would cut for a signal traveling vertically upward with the
speed of light, $c$.
Setting $a$ to $1$ nsec/m, the acceptance cone around
the opposite zenith is slightly increased.
%where $dv=1$m/nsec and $\delta=10$ns is the parameter that
%takes into account the instrumental uncertainty in the time
% measurement.
This condition has been used for almost vertically
up-going muons selection earlier \cite{APP2,JF}.

8608 events survive the selection criterion (4).
Fig.6 shows the hit multiplicity distribution
for these events (dashed histogram)
as well as the expected one for background showers
produced by atmospheric muons (solid histogram).
The experimental distribution is consistent with the theoretical
expectation within a factor 2.
This difference can be explained by the uncertainty
of the atmospheric muon flux close to horizon at the detector
\mbox{depth \cite{APP},} and by uncertainties in the dead-time of
individual channels.
The highest multiplicity of hit
channels (one event) is $N_{hit}=24$.

Since no events with $N_{hit}>24$ are found in our data we can derive
upper limits on the flux of high energy neutrinos which produce
events with multiplicity

\begin{equation}
N_{hit}>25.
\end{equation}
The effective volume of
NT-96 for neutrino induced events depends only slightly on the value of the
threshold multiplicity
in condition (5).
For the stronger conditions
$N_{hit}>27$ and $N_{hit}>29$, the  effective volume decreases
by only 11\% and 27\%, respectively.

The effective volume for neutrino produced events which fulfill
conditions (4)-(5) was calculated as a function of neutrino energy
and zenith angle $\theta$.

The energy dependences of the effective volumes for isotropic electron
and muon neutrinos
are shown in Fig.7 (solid lines).
Also shown are the effective volumes folded with the neutrino
absorption probability in the Earth (dashed lines).
The neutrino absorption in the Earth
has been taken into account with suppression factor
$\exp(-l(\Omega)/l_{tot})$, where $l(\Omega)$ is the neutrino
path length through the Earth in direction $\Omega$ and
$l^{-1}_{tot}=N_A \, \rho_{Earth} \, (\sigma_{CC}+\sigma_{NC})$
according \mbox{to \cite{Gandi,Ber2}.}

\subsection{The limits to the diffuse neutrino flux}

The number of events due to neutrino flux $\Phi_{\nu}$ and processes (1) and
(2) is given by

\begin{equation}
N_{\nu}=T\epsilon \int d\Omega \sum_{\nu}
\int dEV_{eff}(\Omega,E)\sum_k
\int
dE_{\nu}\Phi_{\nu}(E_{\nu}) N_{A} \rho_{H_2O} \frac{d\sigma_{\nu k}}{dE}
\end{equation}
where $E_{\nu}$ is the neutrino energy and $E$ - the energy
transferred to a shower or high energy muon.
The index $\nu$ indicates the summation over neutrino types
($\nu=\nu_{\mu},\bar{\nu_{\mu}},\nu_e$) and
$k$ over CC and NC interactions.
$N_{A}$ is the Avogadro number,
$\epsilon=0.9$ - the detector efficiency.
%the efficiency which includes
%a factor 0.7 taking into account the reduction of the
%effective volume due to light scattering in water (which was not
%included in the MC simulation)
%and the detector efficiency $0.9$.
%The last factor takes into account that the MC calculation
%has been performed for the fully operational NT-96,
%which is larger than the actual configurations
%during 70 days exposition.
Cross sections
%$\frac{d\sigma_{\nu k}}{dE}$
from R.Gandhi {\it et al.}(1996)
have been used.

The shape of the neutrino spectrum was assumed to behave like
$E^{-2}$ as typically expected for Fermi acceleration.
In this case, 90\% of expected events would be produced by  neutrinos
from the energy
range $10^4 \div 10^7$GeV with the center of gravity around $2 \times 10^5$GeV.
Comparing the calculated rates with the upper limit to the
actual number of
events, 2.3 for 90\% CL, and
assuming the flavor ratios $\Phi_{\nu_{\mu}}=\Phi_{\bar{\nu_{\mu}}}=
\Phi_{\nu_e}$ due to photo-meson production of $\pi^+$ followed by
the decay $\pi^+ \rightarrow \mu^+ + \nu_{\mu} \rightarrow
e^+ + \nu_e + \bar{\nu_{\mu}} +\nu_{\mu}$
for extraterrestrial sources \cite{BAHC,P98},
we obtain the following upper
limit to the diffuse neutrino flux:

\begin{equation}
\frac{d\Phi_{\nu}}{dE}E^2<1.43\times10^{-5}
\mbox{cm}^{-2}\mbox{s}^{-1}\mbox{sr}^{-1}\mbox{GeV}.
\end{equation}

New theoretical upper bounds to the intensity of high-energy
neutrinos from extraterrestrial sources have been presented
recently \cite{BAHC,P98}.
These upper bounds as well as our limit and limits obtained
by DUMAND \cite{DUMAND}, AMANDA \cite{AMANDA}, EAS-TOP \cite{EAS}
and FREJUS \cite{FREJUS} experiments are shown in Fig.8.
Also, the atmospheric neutrino \mbox{fluxes \cite{LIP}}
from horizontal  and vertical directions (upper and lower curves,
respectively)
are presented.

For resonant process (3) the event number is given by:

\begin{equation}
N_{\bar{\nu_e}}=T\epsilon \int d\Omega
\int dEV_{eff}(\Omega,E)
\int\limits_{(M_w-2\Gamma_w)^2/2m_e}^{(M_w+2\Gamma_w)^2/2m_e}
%{E_0-\DeltaE}^{E_0+\DeltaE}
dE_{\nu}\Phi_{\bar{\nu_e}}(E_{\nu})
\frac{10}{18}N_{A} \rho_{H_2O} \frac{d\sigma_{\bar{\nu_e},e}}{dE}
\end{equation}

$$
M_w=80.22 \mbox{GeV}, \, \, \, \Gamma_w=2.08 \mbox{GeV}.
$$

Our 90\% CL limit at the W resonance energy is:

\begin{equation}
\frac{d\Phi_{\bar{\nu}}}{dE_{\bar{\nu}}} \leq 3.6 \times
10^{-18}
\mbox{cm}^{-2}\mbox{s}^{-1}\mbox{sr}^{-1}\mbox{GeV}^{-1}.
%cm^{-2}s^{-1}sr^{-1}GeV^{-1}.
\end{equation}
This limit lies between  limits obtained by DUMAND ($1.1 \times 10^{-18}$
cm$^{-2}$s$^{-1}$sr$^{-1}$GeV$^{-1}$) and EAS-TOP
($7.6 \times 10^{-18}$cm$^{-2}$s$^{-1}$sr$^{-1}$GeV$^{-1}$).

The new limits (10) and (12) have been obtained with the underwater
detector NT-96.
%The limit (10) obtained for the diffuse neutrino flux slightly improves
%(and confirms) the
%limit presented by Frejus five years ago \cite{FREJUS}.
We hope that the analysis of 3 years
data taking with NT-200 \cite{APP2,JF} will allow us to lower
this limit substantially.

\subsection{Search for Fast Monopoles ($\beta > 0.75$)}

      Fast bare monopoles with unit magnetic Dirac charge and
 velocities greater than the  Cherenkov  threshold  in  water
 ($\beta = v/c > 0.75$)  are  promising  survey  objects  for
 underwater neutrino telescopes. For a given velocity $\beta$
 the  monopole  Cherenkov  radiation  exceeds   that   of   a
 relativistic  muon  by  a   factor   $(gn/e)^2=8.3\times10^3$
 ($n=1.33$ - index of  refraction  for  water)  \cite{Fr,DA}.
 Therefore  fast  monopoles  with  $\beta  \ge  0.8$  can  be
 detected  up  to  distances  $55$  m  $\div$  $85$  m  which
 corresponds to effective areas of (1--3)$\times 10^4$ m$^2$.

      The natural way for fast monopole detection is based on
 the selection of events with high multiplicity of  hits.  In
 order to reduce the  background  from  downward  atmospheric
 muons we restrict ourself to monopoles coming from the lower
 hemisphere.

Two independent approaches have been used for selection of upward monopole
candidates from the 70 days of NT-96 data.
The first one is similar to the method which was applied
to upward moving muons (see subsection 3.2), with an additional cut
$N_{hit}>25$ on the hit multiplicity.
The second one cuts on the value of space-time correlation, followed
by a cut $N_{hit}>35$ on the hit multiplicity.
The upper limits on the monopole flux
obtained with the two different methods coincide
within errors.

The same type of analysis was applied to
the data taken during $0.42$ years lifetime with the neutrino
telescope NT-36 \cite{INR}.

The combined $90\%$ C.L. upper limit obtained by the Baikal
experiment for an isotropic flux of bare fast magnetic monopoles
is shown in Fig.9, together with the best limits from
underground experiments Soudan2, KGF, MACRO and Ohya
\cite{Oh,MA,KGF,Sou}.

\section{Limnology}

      Besides physics goals NT-200 can be used as a  powerful
 tool to monitor water  parameters.  The  array  permanently
 records  phototubes   counting   rates,   and   periodically
 parameters like optical transmission at various wavelengths,
 temperature, conductivity, pressure, and sound velocity. All
 these data complement  traditional limnological
 studies  are  of   importance   to   get   a   comprehensive
 understanding of processes occurring in the lake.
      Just  for  illustration  we  show  OMs  counting   rate
 variations at various time scales  recorded  with  NT-36  in
 1993/94.  Counting  rates  of  individual  OMs  as  well  as
 individual channels (coincidence rate  of  a  pair  OMs)  are
 dominated by water luminescence.
      Fig.10 presents the counting rate  over  2  years  and
 compares  it  to  the  bacteria  concentration  measured  at
 distance of \mbox{50 km} to the NT-200 site, at \mbox{10 m}  depth  below
 surface. In August/September we observe an increase  of  the
 luminosity to extremely high  levels.  The  changes  of  the
 counting rate of channels are  not  reflected  in  the  muon
 trigger  rate,  since  the  muon  trigger  is  dominated  by
 atmospheric muons, with negligible  contribution  by  random
 hits (water luminescence or dark current  pulses).  This  is
 demonstrated in fig.11 on a shorter time scale, for a  time
 interval of marked changes of individual channel counting
 rates following a severe storm at August  3rd,  1993,  which
 had washed a lot of water from a nearby rivers and crooks to
 the lake.
      Fig.12 shows a short period of about 8 hours  when  the
 counting rates sequentially increased, starting with highest
 OMs and ending with the lowest ones along the  string.  From
 the time shift of the 3 curves a  vertical  current  of  $2.3$ cm$\cdot$s$^{-1}$
 is deduced. This is remarkable since the vertical
 velocity  of  water  renewal  is  considered  to   be   most
 intensively, is only $0.2\div0.3$ cm$\cdot$s$^{-1}$.

\section{Conclusions and Outlook}
The results obtained with intermediate detector stages
show the capability of Baikal Neutrino Telescope
to search for the wide variety of phenomena
in neutrino
astrophysics, cosmic ray physics and particle physics.

%The Baikal detector is well understood, and
The first atmospheric neutrinos have
been identified. Also limits on the fluxes of magnetic monopoles, 
diffuse flux of very high energy neutrinos as well as
of neutrinos from WIMP annihilation in the center of the Earth
have been derived.

In the following years, NT-200 will be operated as a
neutrino telescope with an effective area between
1000 and 5000 m$^2$, depending on the energy, and
will investigate atmospheric neutrino spectra above 10 GeV.
%(about 1$\div$2 atmospheric neutrinos per 2 days).

NT-200 can be used to
search for neutrinos from WIMP annihilation and for
magnetic monopoles and high energy extraterrestrial neutrinos. It will also be a unique
environmental laboratory to study water processes
in Lake Baikal.

Apart from its own goals, NT-200 is regarded to be a
prototype
for the development a large scale telescope of next generation
with an effective area of
several $10^4$ m$^2$.
%This telescope would have a
%realistic detection potential for extraterrestrial sources
%of high energy neutrinos.
The basic design of such a detector is under
discussion at present.

\bigskip

{\it This work was supported by the Russian Ministry of Research,the German
Ministry of Education and Research and the Russian Fund of Fundamental
Research \mbox{( grants} } \mbox{\sf 99-02-18373a}, \mbox{\sf 97-02-17935},
\mbox{\sf 99-02-31006k}, \mbox{\sf 97-02-96589} {\it and} \mbox{\sf 97-05-96466}).
%\newpage

\newpage

\bigskip

\begin{table}
\tcaption{\normalsize
The expected number of atmospheric neutrino events and background
events, and the observed number of events after cuts 1--6.}
%\small
\vspace{0.5cm}
\begin{center}
\begin{tabular} {||c|c|c|c|c|c|c||} \hline \hline
after cut {\cal N}$^o$  $\rightarrow$ & 1 & 2 & 3 & 4 & 5 & 6 \\ \hline %\hline
atm. $\nu$, MC & 11.2 & 5.5 & 4.9 & 4.1 & 3.8 & 3.5 \\ \hline
background, MC & 7106 & 56 & 41 & 16 & 1.1 & 0.2 \\ \hline
experiment     & 8608 & 87 & 66 & 28 & 5 & 4 \\ \hline \hline
\end{tabular}
\end{center}
\end{table}

\bigskip

\begin{table}[ht]
\label{limit}
%\footnotesize\rm
\tcaption{\normalsize 90\% C.L. upper limits on the muon flux from the center of the Earth
for four regions of zenith angles obtained in different experiments
}
%\small
\vspace{0.5cm}
\begin{center}
\begin{tabular}{||c|c|c|c|c||} \hline \hline
      &\multicolumn{4}{c|}{Flux limit       ($10^{-14} \cdot (cm^2 \; sec)^{-1})$} \\ \cline{2-5}
Zenith   & NT-96  & {\it Baksan} & {\it MACRO}  & {\it Kam-de}  \\
angles   &$>10GeV$       &$>1GeV$        &$>1.5GeV$      &$>3GeV$ \\ \hline
$\geq 150^{\circ}$  & $11.0$      & $2.1$ & $2.67$ &$4.0$ \\ \hline
$\geq 155^{\circ}$  & $9.3 $      & $3.2$ & $2.14$ &$4.8$ \\ \hline
$\geq 160^{\circ}$  & $ 5.9-7.7 $ & $2.4$ & $1.72$ &$3.4$ \\ \hline
$\geq 165^{\circ}$  & $4.8$       & $1.6$ & $1.44$ &$3.3$ \\ \hline \hline
\end{tabular}
\end{center}
\end{table}

\newpage

\begin{figure}[p]
\centering
\mbox{\epsfig{file=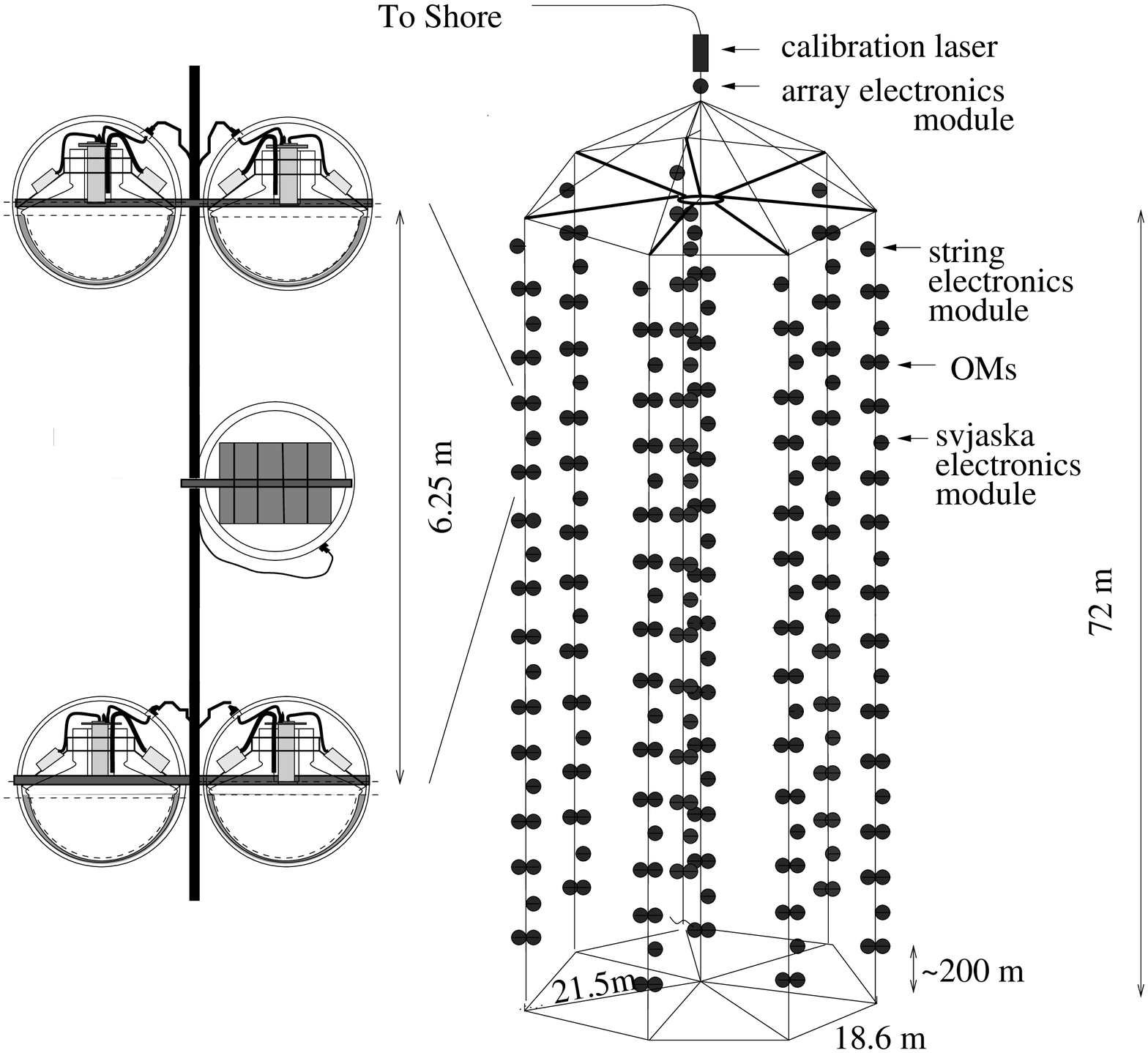,width=15cm}}
  {
    \fcaption{\small
       Schematic view of the Baikal Telescope NT-200. The array
       is time-calibrated by two nitrogen lasers. The one (fiber laser)
       is mounted just above the array. Its light is guided via optical
       fibers to each OM pair. The other (water laser) is arranged
       20\,m below the array. Its light propagates directly through the
       water. The expansion left-hand shows 2 pairs of
      optical modules ("svjaska") with the svjaska
      electronics module, which houses
      parts of the read-out and control electronics.
}}
\end{figure}

\newpage

\begin{figure}[p]
\centering
\mbox{\epsfig{file=complex.eps,width=15cm}}
  {
    \fcaption{
    The principle of joint operation of NT-200 with an EAS Cherenkov
    array. Also shown (bottom right) are optical modules contained in a light tight tank.
    Their signal fixes one track point as a tool for calibration.
  }
}
\end{figure}

\begin{figure}[p]
\centering
\mbox{\epsfig{file=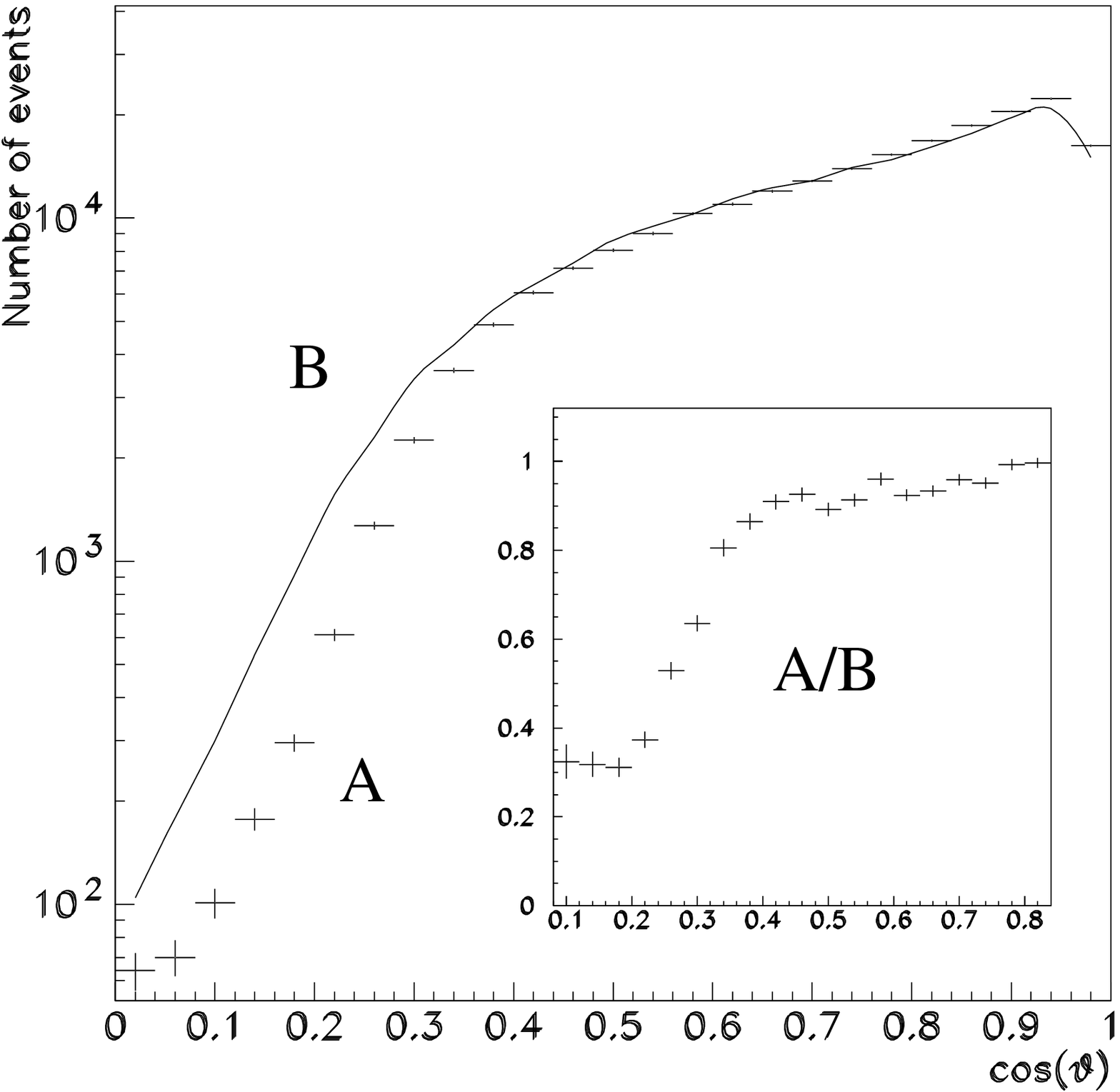,width=15cm}}
{\fcaption {
Atmospheric muons (vs zenith angle $\theta$) as it is
measured in the direction to the nearest
point of the shore(A) and in the opposite direction (B) - ``open'' water.
The small picture shows the ratio A to B.
}}

\end{figure}

\newpage

\begin{figure}
\centering
\mbox{\epsfig{file=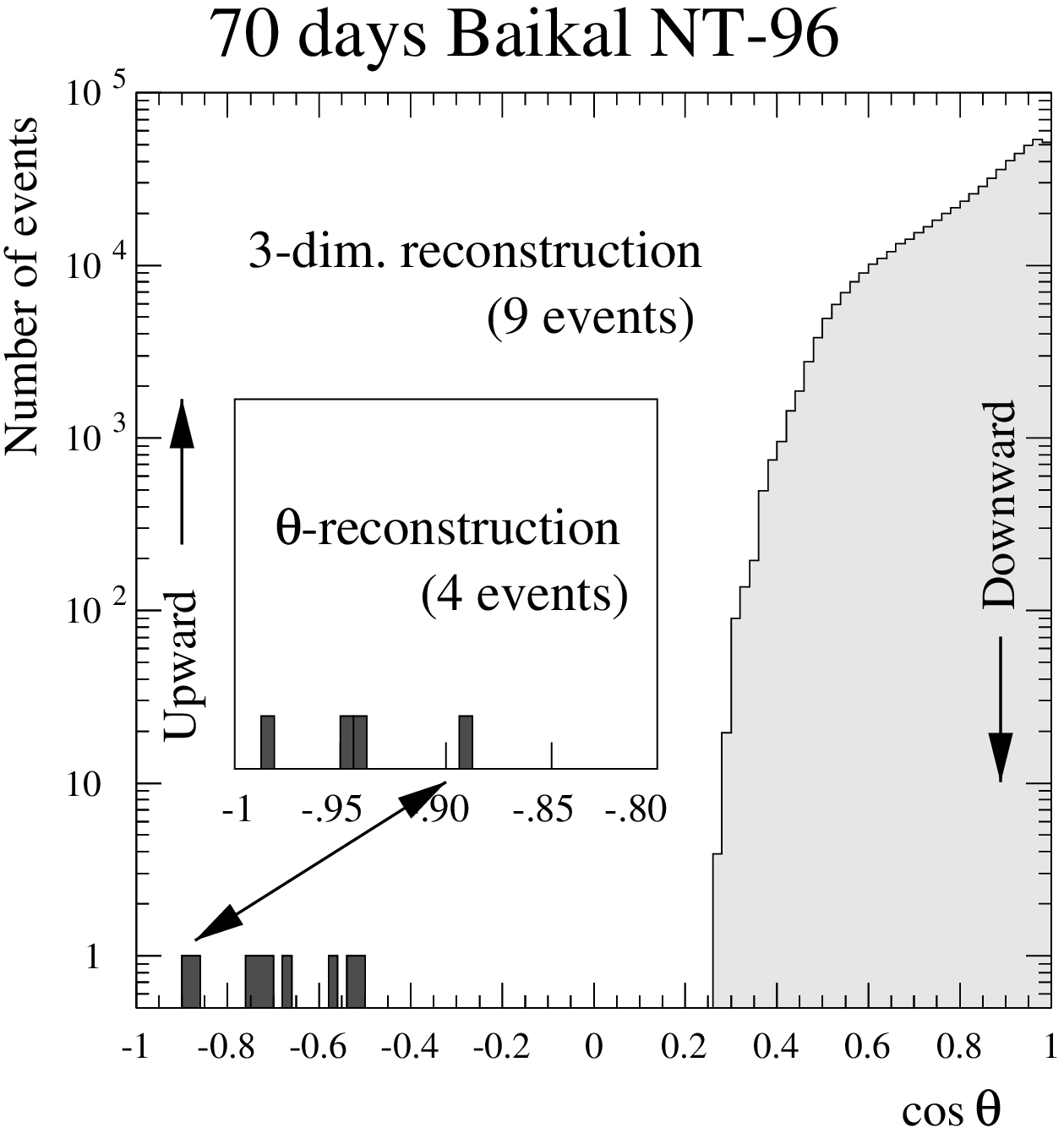,width=15cm}}
\fcaption{\small
Experimental angular distribution of events satisfying
trigger
{\it 9/3}, all final quality cuts and the limit on $Z_{dist}$ (see
text). The small picture shows the events selected by using the method 
described in subsection 3.1. The event found by both algorithms is
marked by the arrow.
}
\end{figure}

\newpage

\begin{figure}
\centering
\mbox{\epsfig{file=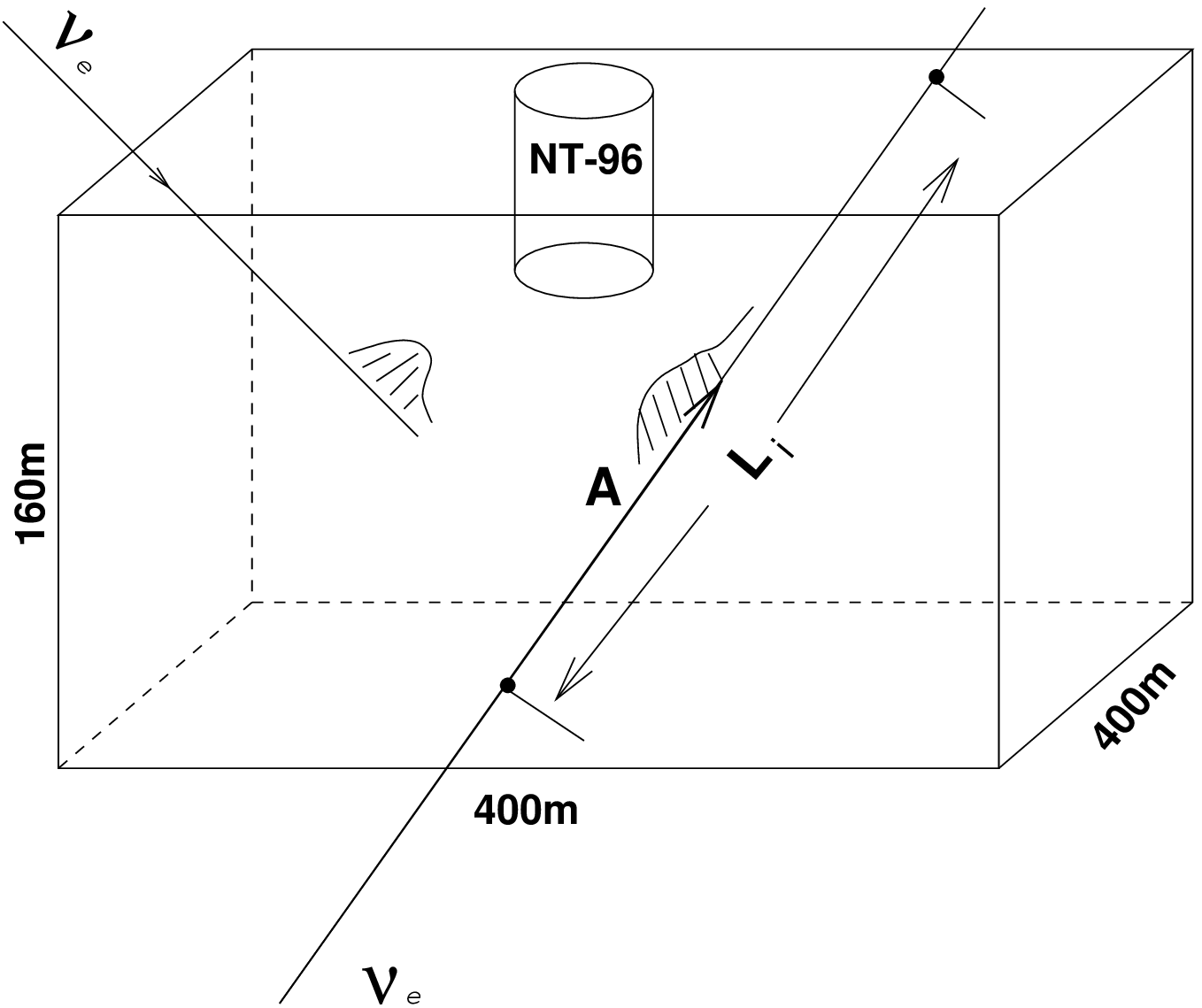,width=15cm}}
{\fcaption { 
Detection principle for neutrino induced
showers with NT-96.
}}
\end{figure}

\newpage

\begin{figure}
\centering
\mbox{\epsfig{file=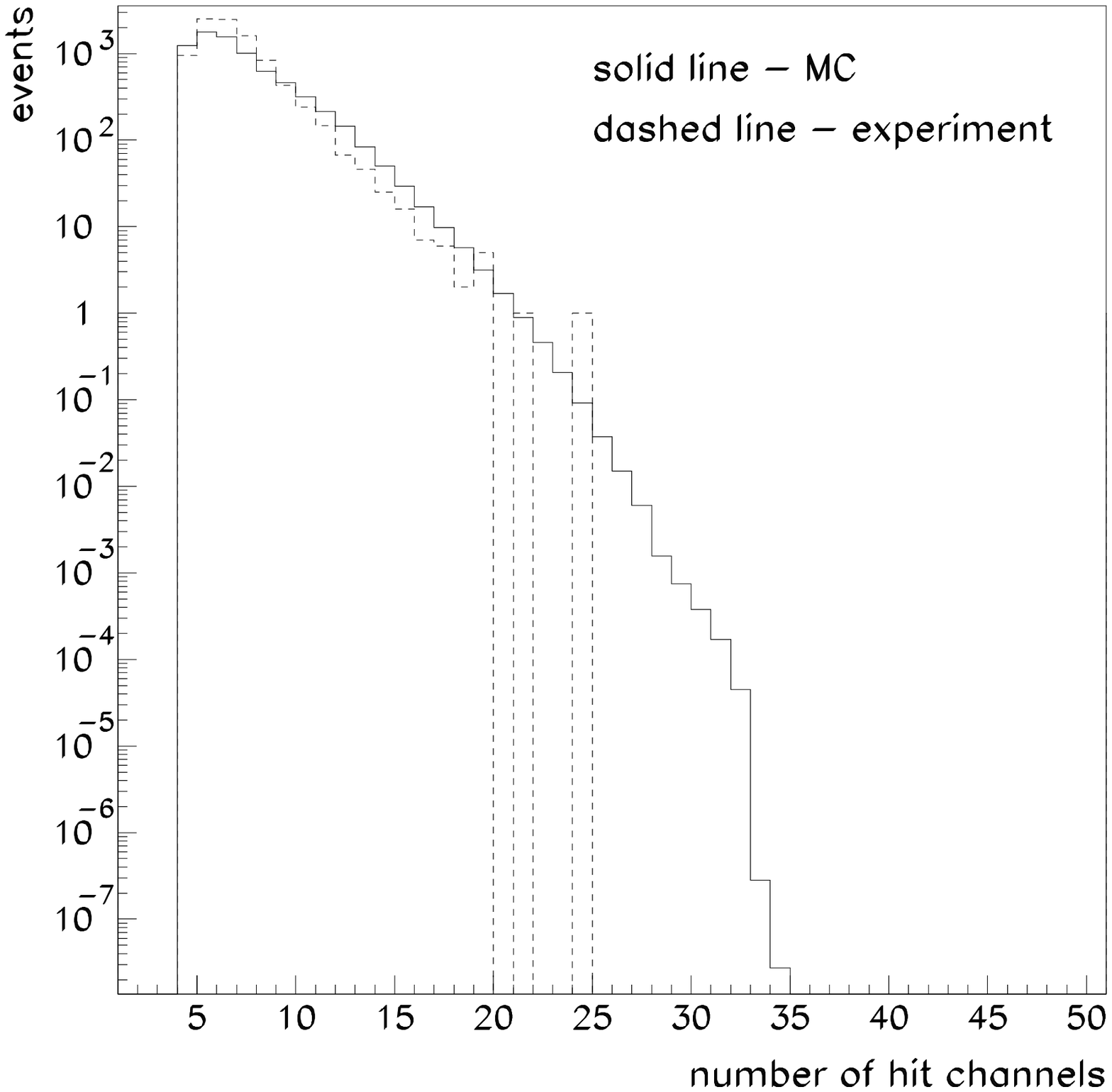,width=15	cm}}
{\fcaption { 
Hit channel multiplicity: solid histogram - showers produced by
atmospheric muons (MC), dashed histogram - experiment.
}}
\end{figure}

\newpage

\begin{figure}
\centering
%\mbox{\epsfig{file=meanven.eps,width=9.7cm}}
\mbox{\epsfig{file=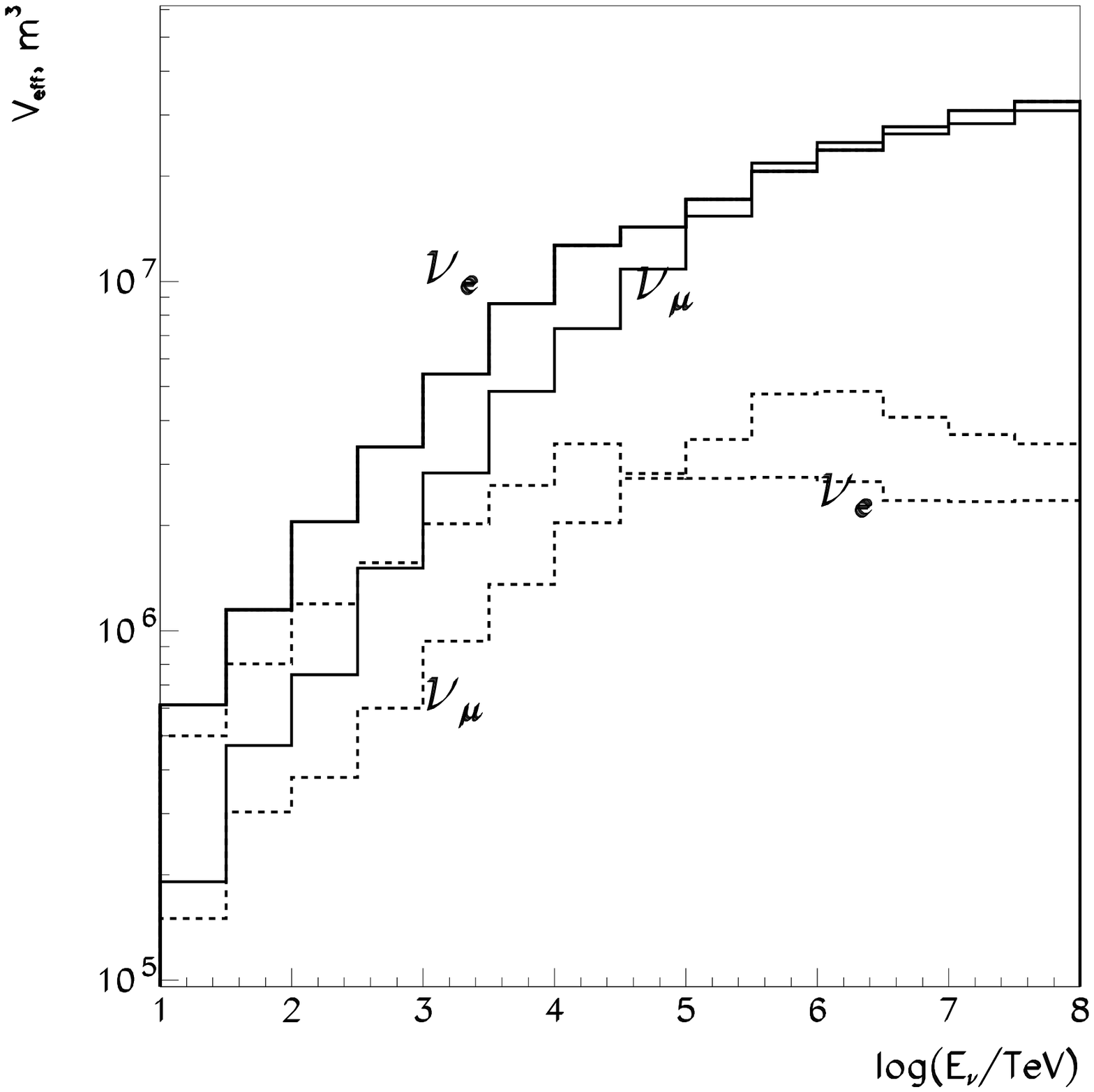,width=15cm}}
{\fcaption { 
Effective volumes of NT-96 for isotropic electron and muon neutrinos
(solid lines). 
The dashed lines represent the effective volumes folded with 
the neutrino 
absorption probability in the Earth (see text).
%$\int d\Omega V(\Omega,E)\exp(-l(\Omega)/l_{tot})$. 

}}
\end{figure}

\newpage

\begin{figure}
\centering
\mbox{\epsfig{file=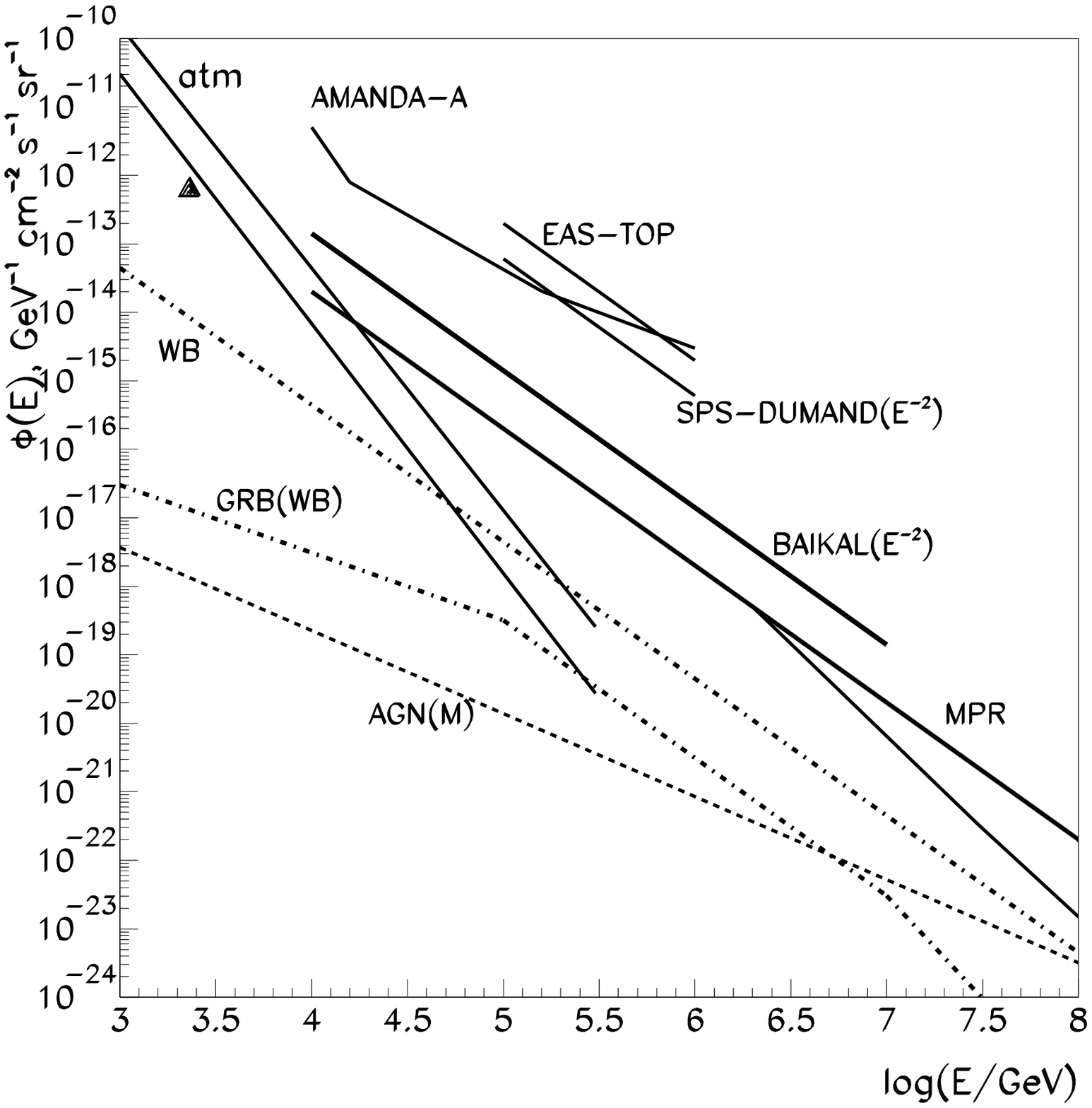,width=12.8cm,height=15cm}}
{\fcaption { 
Upper limits to the differential flux of high 
energy neutrinos obtained by 
different experiments as well as upper bounds for  
neutrino fluxes from a number of different models. 
Dot-dash curves
labeled WB and GRB(WB) - upper bound and neutrino intensity
from GRB estimated by Waxman and Bahcall (1997,1999);
dashed curve labeled AGN(M) - neutrino intensity from AGN (Mannheim 
model A, 1996);
solid curves labeled MPR - upper bounds for $\nu_{\mu}+\bar{\nu_{\mu}}$
in Mannheim et al. (1998) for pion photo-production neutrino sources
with different optical depth $\tau$ (adapted from ref.17).  
The triangle denotes the limit obtained by the Frejus-Experiment
for an energy of \mbox{2.6 TeV :}  $7 \times 10^{-13} \mbox{cm}^{-2}
\mbox{s}^{-1} \mbox{sr}^{-1} \mbox{GeV}^{-1}$.
}}
\end{figure}

\newpage

\begin{figure}
\centering
\mbox{\epsfig{file=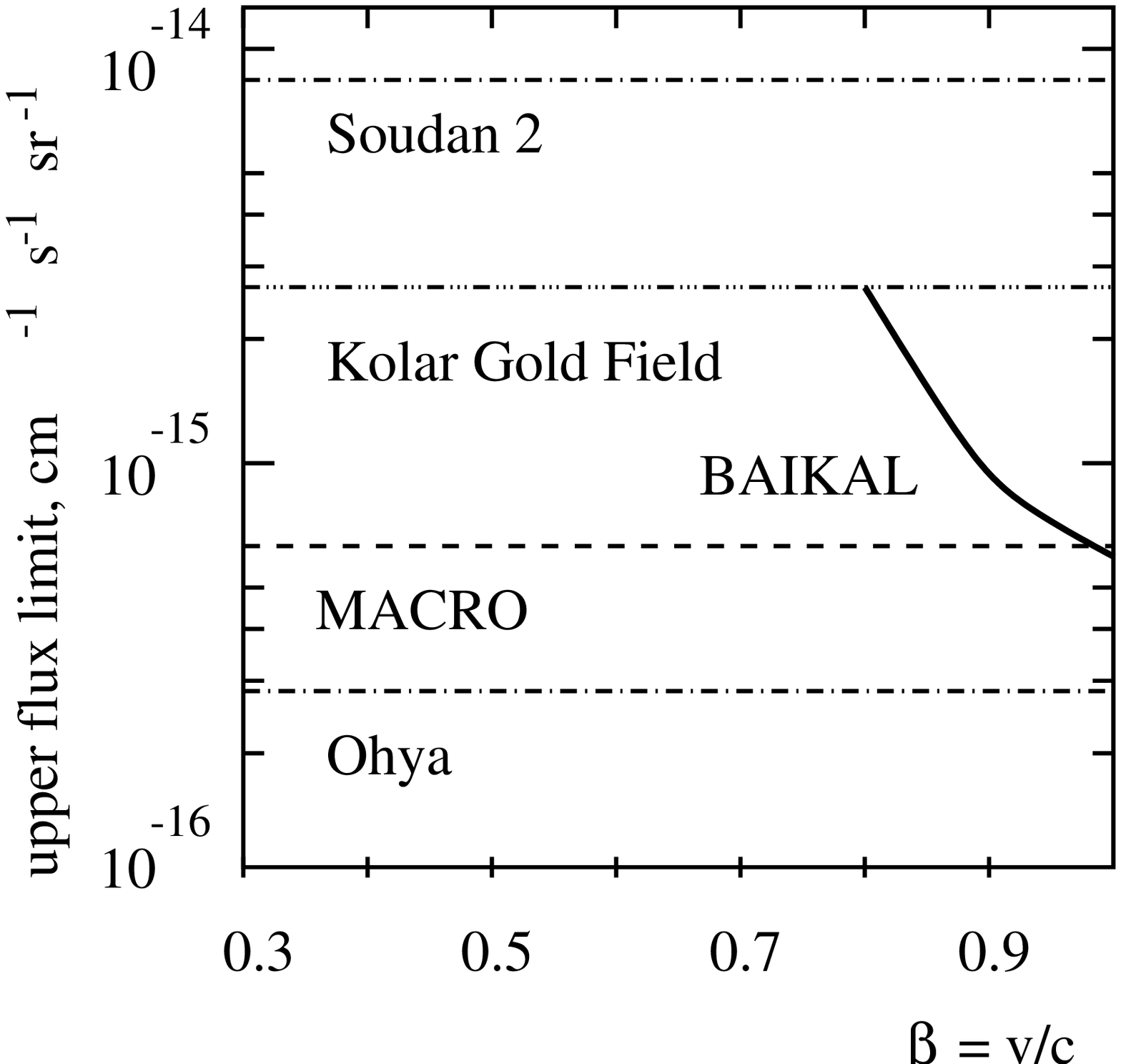,width=15cm}}
{\fcaption { 
The $90\%$ C.L. Baikal upper limit for an isotropic flux of
bare magnetic monopoles compared with other published limits.   
}}
\end{figure}

\newpage

\begin{figure}[p]
\centering
\mbox{\epsfig{file=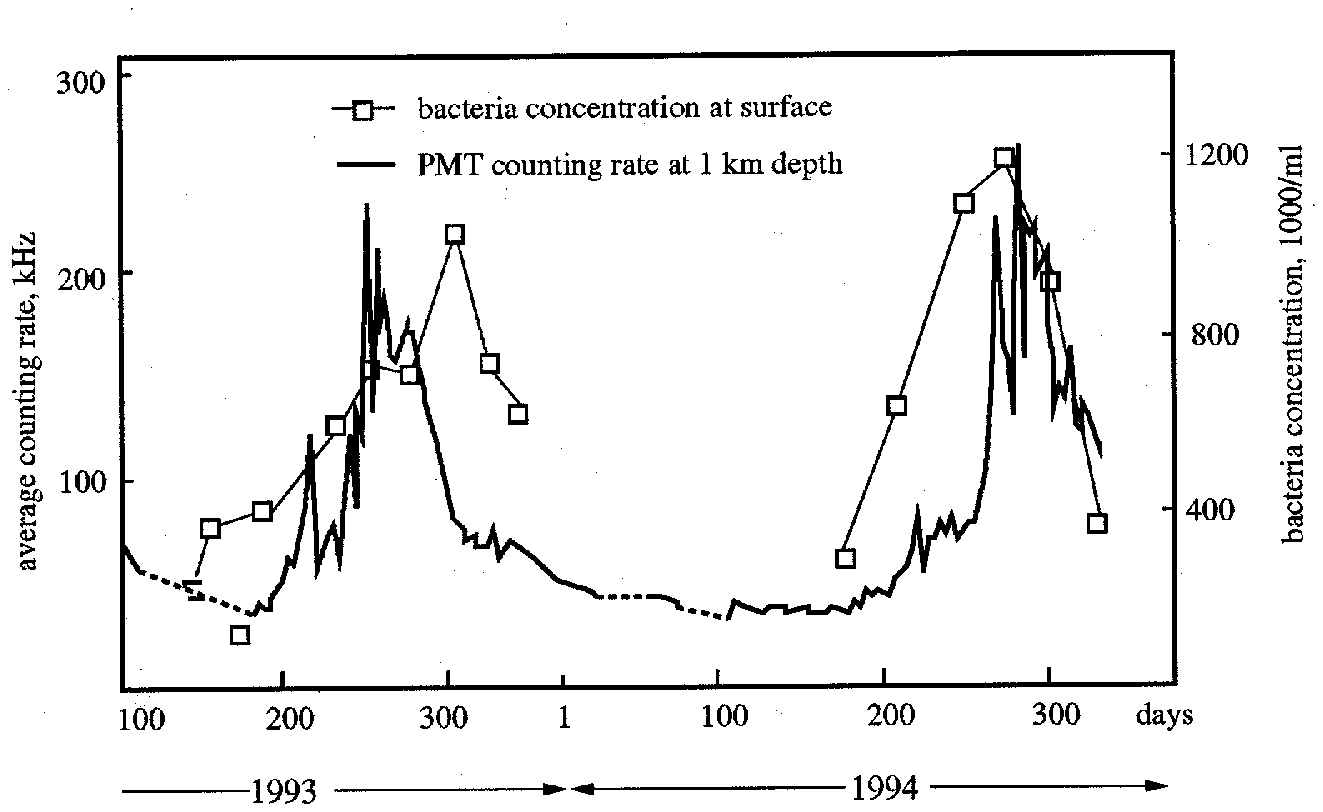,width=15cm}}
  {
    \fcaption{
    Average counting rate of OMs vs. time, compared to bacteria concentration at surface.
  }
}
\end{figure}

\newpage

\begin{figure}[p]
\centering
\mbox{\epsfig{file=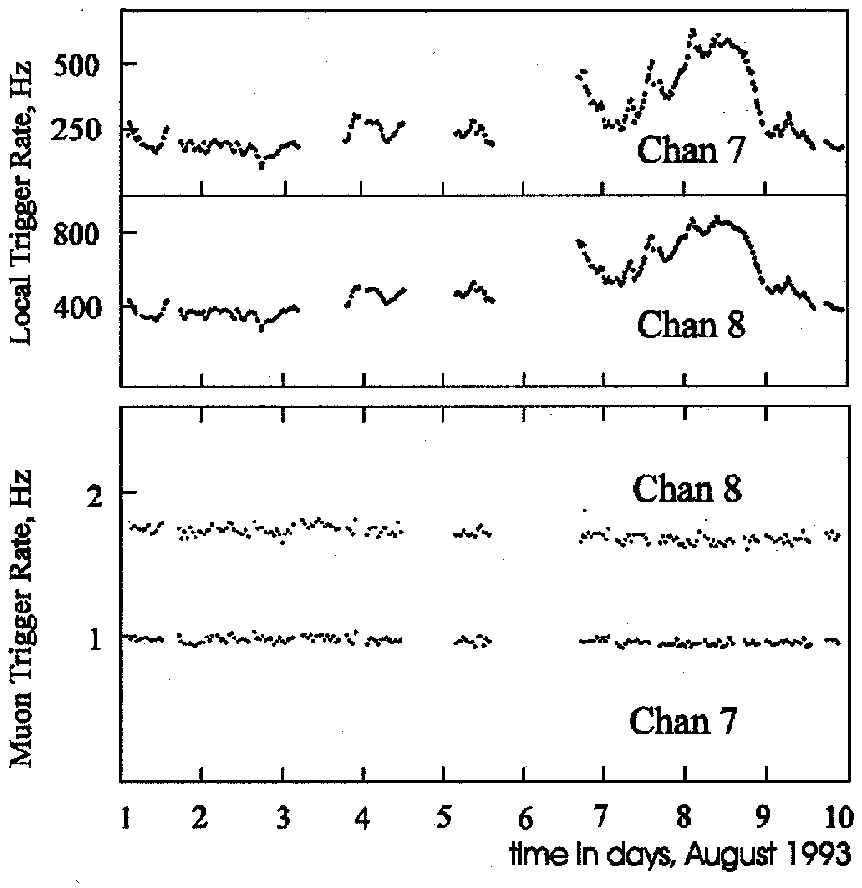,width=15cm}}
  {
    \fcaption{
    a) Local trigger rates for channel 7 (downward facing) and channel 8 (upward facing)
    for August 1st-9th, 1993. The counting rates are averaged over 30 min. b) Mouns trigger rates 
    (condition 4/1) for channel 7 and 8. Counting rates are averaged over 50 min.
  }
}
\end{figure}

\newpage

\begin{figure}[p]
\centering
\mbox{\epsfig{file=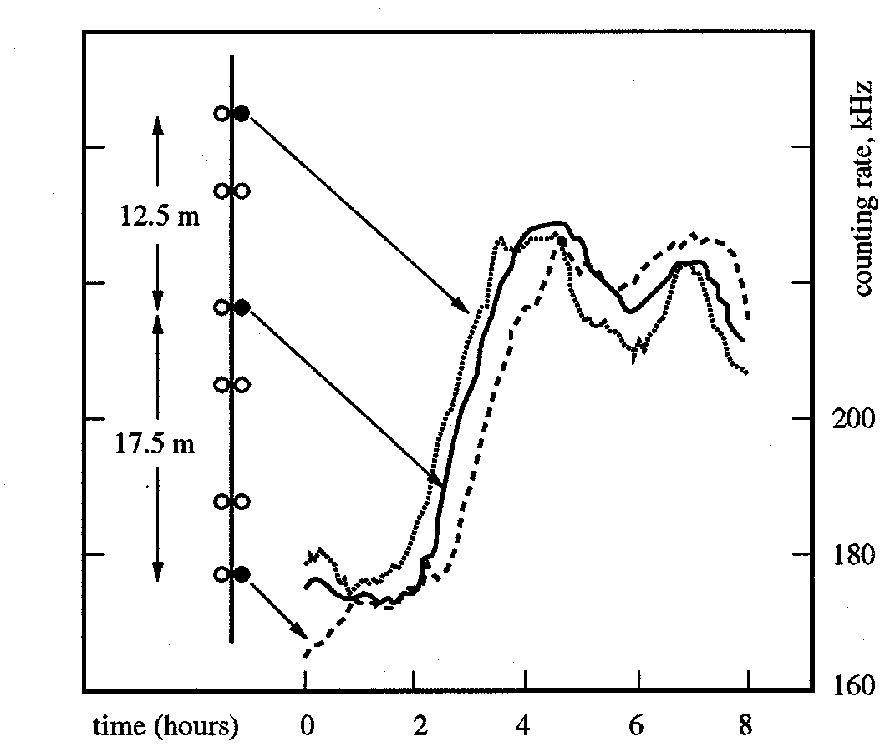,width=15cm}}
  {
    \fcaption{
    Counting rate of three OMs along one string during an 8 hour interval at Sept. 24, 1993.
  }
}
\end{figure}


\vspace{1cm}
{\bf REFERENCES}

\begin{thebibliography}{99}

\bibitem{Project}
{\it The Baikal Neutrino Telescope NT-200, BAIKAL 92-03} (1992)

\bibitem{APP} I.A.Belolaptikov {\it et al.},
{\it Astroparticle Physics} 7 (1997) 263.

\bibitem{APP2} I.A.Belolaptikov {\it et al.},
{\it astro-ph/9903341} (1999) (accepted for publ. in {\it Astropart. Phys.}).


\bibitem{NIM} R.I.Bagduev {\it et al.,} {\it Nucl. Instr. Meth.}
{\bf A420} (1999) $138\div154$.
\bibitem{Aachen} R.I.Bagduev {\it et al.}, {\em Proc. Int. Conference on
                 Trends in Astroparticle Physics}, 132 (Aachen, 1994)

\bibitem{Pylos} L.B.Bezrukov {\it et al.}, {\em Proc. 3rd NESTOR Workshop,
                Oct. 1993, Greece, $645\div657$.}
		
\bibitem{Calg} R.A.Antonov {\it et al.,} {\it Proc. 23-rd ICRC} Calgary, vol.2 (1993) $430\div433$

\bibitem{FRST_vert} L.B.Bezrukov {\it et al.},
{\em Proc. of the 2nd Workshop on the
Dark Side of the Universe},
221 (Rome, 1995) ({\sf astro-ph/9601161})

\bibitem{INR_vert} V.A.Balkanov {\it et al.} 1998 {\it Preprint INR
0972/98} (in russian)

\bibitem{Bak} M.M.Boliev {\it et al.}  {\em Nucl.Phys.} (Proc. Suppl.) {\bf 48} (1996) 83

\bibitem{MACRO} T.Montaruli {\it et al.}  {\it Proc. 25-th ICRC} Durban--South Africa, vol.7, (1997) 185

\bibitem{Kam} M.Mori {\it et al.}  {\em Phys. Rev.} {\bf D48} (1993) 5505

\bibitem{Mosc} L.Moscoso {Oral presentation at NOW98} 1998.

\bibitem{DUMAND} J.W.Bolesta {\it et al.} {\it Proc. 25-th ICRC} Durban--South Africa, vol.7, (1997) 29

\bibitem{AMANDA} R.Porrata {\it et al.} {\it Proc. 25-th ICRC} Durban--South Africa, vol.7, (1997) 9

\bibitem{Glash} S.L.Glashow, {\it Phys. Rev.} {bf 118} (1960) 316

\bibitem{Ber1} V.S.Beresinsky and A.Z.Gazizov, {\it JETP Lett.} {\bf 25} (1977) 254

\bibitem{Gandi} R.Gandhi {\it et al.}  {\it Astroparticle Physics} {\bf 5} (1996) 81

\bibitem{JF} V.A.Balkanov {\it et al., Physics of Atomic Nuclei} {\bf 62} (1999) 949

\bibitem{Ber2} V.S.Berezinsky {\it et al., Sov. J. Nucl. Phys.} {\bf 43} (1986) 406

\bibitem{BAHC} E.Waxman, and J.Bahcall, {\it Phys. Rev.} {\bf D 59} (1999) 023002; E.Waxman, and J.Bahcall, 
{\it Phys. Rev. Lett.} {\bf 78} (1997) 2292

\bibitem{P98} R.J.Protheroe   {\it e-preprint astro-ph/9809144}  (1998)

\bibitem{EAS} M.Aglietta {\it et al.}  {\it Phys. Lett.} {\bf B333} (1994) 555


\bibitem{FREJUS} W.Rhode {\it et al., Astropart. Phys.} {\bf 4} (1994) 217

\bibitem{LIP} P.Lipari, {\it Astropart. Phys.} {\bf 1} (1993) 195

\bibitem{Fr} I.M.Frank 1988 {\it Vavilov-Cherenkov Radiation} (Moscow:
Nauka) 192 (in russian)

\bibitem{DA} D.A.Kirzhnits and V.V.Losjakov  {\it Pis'ma Zh. Eksp.Theor.Fz.} {\bf 42} (1985) 226

\bibitem{INR} V.A.Balkanov {\it et al.}  {Preprint INR} (Moscow: INR) (1998) (in russian)


\bibitem{Oh} S.Orito {\it et al.}  {\it Phys. Rev. Lett.} {\bf 66} (1992) 1951.

\bibitem{MA} M.Ambrosio {\it et al.}  {\it MACRO Preprint} MACRO/PUB 98/3 (1998)

\bibitem{KGF} H.Adarkar {\it et al.}  {\it Proc. 21st ICRC. Adelaide.}  (1990) 95

\bibitem{Sou} J.L.Thorn {\it et al.}  {\it Phys. Rev.} {\bf D46} (1992) 4846

\end{thebibliography}
\end{document}